\documentstyle[12pt,epsfig]{article}                                         
                                         
\newlength{\dinwidth}                                         
\newlength{\dinmargin}                                         
\def\lapproxeq{\lower .7ex\hbox{$\;\stackrel{\textstyle                                         
<}{\sim}\;$}}                                         
\def\gapproxeq{\lower .7ex\hbox{$\;\stackrel{\textstyle                                         
>}{\sim}\;$}}                                         
\def\be{\begin{equation}}                                         
\def\ee{\end{equation}}                                         
\def\bea{\begin{eqnarray}}                                         
\def\eea{\end{eqnarray}}                                         
                                         
\def\funp{{I\!\!P}}

\begin{document}                                         
\titlepage                                         
\begin{flushright}                                         
DTP/99/98 \\
hep-ph/9912551 \\                                         
December 1999 \\                                         
\end{flushright}                                         
                                         
\vspace*{2cm}                                         
                                         
\begin{center}                                         
{\Large \bf $Q^2$ dependence of diffractive vector meson electroproduction}                                         
                                         
\vspace*{1cm}                                         
A.D.~Martin$^a$, M.G.~Ryskin$^{a,b}$ and T.~Teubner$^c$ \\                                         
                                        
\vspace*{0.5cm}                                         
$^a$ Department of Physics, University of Durham, Durham, DH1 3LE, England\\                                        
$^b$ Petersburg Nuclear Physics Institute, Gatchina, St.~Petersburg, 188350, Russia \\         
$^c$ Institut f\"{u}r Theoretische Physik E, RWTH Aachen, D-52056
Aachen, Germany       
\end{center}                                         
                                         
\vspace*{2cm}                                         
                                         
\begin{abstract}                                         
We give a general formula for the cross section for diffractive vector meson       
electroproduction, $\gamma^* p \rightarrow Vp$.  We first calculate diffractive $q\bar{q}$       
production, and then use parton-hadron duality by projecting out the $J^P = 1^-$ state in the       
appropriate mass interval.  We compare the $Q^2$ dependence of the cross section for the       
diffractive production of $\rho$ and $J/\psi$ mesons with recent HERA data.  We include the      
characteristic $Q^2$ dependence associated with the use of the skewed gluon       
distribution.  We give predictions for $\sigma_L/\sigma_T$ for both $\rho$ and $J/\psi$       
production.      
\end{abstract}                                        
                                
\newpage                                        
The diffractive leptoproduction of vector mesons at high energy is an interesting and          
important process. Indeed diffractive $\gamma^* p \rightarrow Vp$ data, with $V = \rho,        
\omega, \phi, J/\psi$ and $\Upsilon$, are becoming available with increasing precision from        
the experiments at HERA \cite{H1}--\cite{Z2}.  They offer the opportunity to study the        
vacuum-exchange singularity as a function of the virtuality $Q^2$ of the incoming photon        
and of the mass $M$ of the produced vector meson. Moreover observation of the vector      
meson decays allows both $\sigma_L$ and $\sigma_T$ to be measured, and even $s$-channel      
helicity conservation to be checked \cite{H2,H3,Z1,Z2}.       
       
Let us first review the description of diffractive electroproduction of $\rho$ mesons; a process        
which has attracted a lot of theoretical          
interest \cite{DL}--\cite{MRT1}.  At first, phenomenological parametrizations          
based on the vector-meson-dominance model and Regge exchanges were used.  Then a          
non-perturbative two-gluon exchange model of the Pomeron was introduced          
\cite{DL}.  For large $Q^2$ however, we would expect that a pure perturbative QCD          
description is applicable.  Such a description for the production of longitudinally          
polarised $\rho$ mesons was given by Brodsky et al.~\cite{BROD}, using the leading          
twist wave function for the $\rho$ meson.  The process is sketched in
Fig.~1.  However for the production of transversely          
polarized $\rho$ mesons, the perturbative QCD approach encounters an infrared          
divergence in the integration over the quark transverse momentum.  This problem can be          
overcome by using parton-hadron duality \cite{MRT1}.  The wave function of the          
$\rho$ meson then never enters explicitly.  The only property that is used is that the          
$\rho$ meson corresponds to the $J^P = 1^-$ projection of \lq open' $q\bar{q}$          
production (with $q = u,d$).  The projection has the effect of curing the infrared          
divergence. 
\begin{figure}[htb]
\vspace{-0.5cm}
\begin{center}
\leavevmode
\epsfxsize=8.0cm
\epsffile[100 450 440 660]{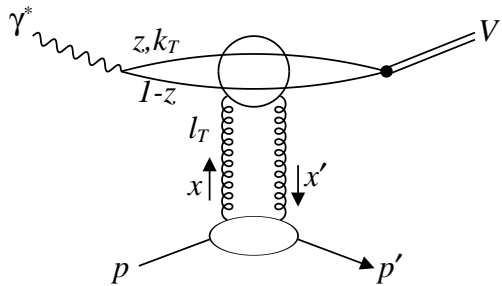}
\end{center}
\vspace{-1.5cm}
\caption[]{\label{fig1} Schematic diagram for diffractive vector meson
  production at HERA, $\gamma^* p \rightarrow Vp$.  The longitudinal
  fractions $x$ and $x^\prime$ of the ingoing and outgoing proton
  momentum carried by the gluons are given by Eq.~(\ref{eq:b9}); the
  gluons have
  momenta $\pm \mbox{\boldmath $\ell$}_T$ transverse to the proton.
  $z$ and $1 - z$ are the longitudinal fractions of the photon
  momentum carried by the $q$ and $\bar{q}$, and $\pm \mbox{\boldmath
    $k$}_T$ are their momenta transverse to the photon.  There are
  four possible couplings of the two gluons to the $q$ and $\bar{q}$,
  represented by the upper circle.}
\end{figure}
The          
resulting cross section is then integrated over an appropriate interval $\Delta M$ of          
the invariant mass of the $q\bar{q}$ pair which covers the $\rho$ resonance peak.  As          
there are almost no other possibilities\footnote{We allow for $\omega$ production by taking       
the ratio $\omega : \rho$ to be 1:9.} for hadronization of the $q\bar{q}$ pairs at          
$M_{q\bar{q}} \simeq M_\rho$, the procedure is expected to give a reasonable          
estimate of the cross section for $\rho$ electroproduction.  Indeed this perturbative          
framework \cite{MRT1} was found to describe the $Q^2$ dependence of $\rho$          
electroproduction for $Q^2 \gapproxeq 5$~GeV$^2$ for both longitudinally and          
transversely polarised $\rho$ mesons, including the observed $Q^2$ dependence of          
the $\sigma_L/\sigma_T$ ratio.  The $Q^2$ behaviour of the amplitude is governed          
by the structure of the quark propagators and by the effective anomalous dimension      
$\gamma$ of the gluon, defined by $xg (x, K^2) \sim (K^2)^\gamma$.  In particular the naive      
expectation that $\sigma_T = \sigma_L M^2/Q^2$ is modified to\footnote{Eq.~(\ref{eq:aa})      
is an approximate result obtained assuming that, for each $Q^2$, $\gamma$ is constant      
throughout the integration over the quark loop.  The full calculation can be found in      
Ref.~\cite{MRT1}.  See also the results discussed below.}         
\be         
\label{eq:aa}         
\frac{\sigma_L}{\sigma_T} \; = \; \frac{Q^2}{M^2} \: \left ( \frac{\gamma}{\gamma          
+ 1} \right )^2         
\ee         
which, on account of the decrease of $\gamma$ with increasing $Q^2$, is in good          
agreement with the observed $\sigma_L/\sigma_T$ behaviour with $Q^2$.         
         
The cross section for the diffractive electroproduction of vector mesons is proportional to the        
square of the off-diagonal or skewed gluon distribution.  That is $x \neq x^\prime$ in Fig.~1,       
whereas for the conventional (diagonal) gluon we have $x = x^\prime$.  In fact the gluon       
distribution becomes more        
skewed as $Q^2$ increases, and is more skewed for vector mesons of larger mass $M$.    
Skewed distributions were not included in our predictions of the $Q^2$ behaviour of $\rho$   
electroproduction in Ref.~\cite{MRT1}.  When compared with subsequent
precise HERA data \cite{H2}, these predictions were found to fall off
a bit too rapidly with increasing $Q^2$.  We will see that the effect
of using the skewed distribution, rather than the usual approximation
of using the conventional (diagonal) gluon, will enhance the cross
sections at the larger values of $Q^2$ at which data exists.  As was
previously discussed~\cite{MRT1}, there are uncertainties in the
normalization of the predictions of the diffractive cross sections,
but much less in the predictions of the energy or $Q^2$ dependence.
Nevertheless, in order to use the $Q^2$ dependence of the data to reveal the
effects of the skewed distribution, we must include the $Q^2$
dependence of other effects in our calculation.  The (imaginary part
of the) amplitude is calculated at $t = 0$ and the cross section
obtained by integrating $d \sigma/dt \sim \exp (- bt)$ over $t$.  We
must therefore allow for the decrease of $b$ with increasing $Q^2$.
Second we must study the ambiguity in our estimates of the
next-to-leading order (NLO) correction.  In the perturbative region,
we find that the $Q^2$ variation of $\rho$ electroproduction from
these two sources is smaller than that due to the use of the skewed
gluon distribution.  Also we must, of course, include the contribution
from the real part of the amplitude.  When we compare the full QCD
prediction with the $Q^2$ behaviour of diffractive $\rho$ and $J/\psi$
production recently measured at HERA we find that the data are
compatible with the characteristic enhancement arising from the skewed
gluon.  
       
We use perturbative QCD to derive the general formula for the cross section for diffractive       
vector meson production by first recalling the formula for diffractive production of a       
$q\bar{q}$ system of mass $M$.  For production from a transversely (longitudinally)       
polarised photon      
\be      
\label{eq:a2}      
\left . \frac{d^2 \sigma^{T (L)}}{dM^2 dt} \right |_{t = 0} \; = \; \frac{2 \pi^2 e_q^2       
\alpha}{3 (Q^2 + M^2)^2} \: \int \: dz \: \left | B_{ii^\prime}^{T (L)} \right |^2      
\ee      
where $i = +, -$ and $i^\prime = +, -$ denote the helicity of the quark and antiquark.  The       
helicity amplitudes are      
\bea      
\label{eq:a3}      
{\rm Im} B_{++}^T & = & \frac{m I_L}{2 \sqrt{z (1 - z)}}, \quad\quad B_{--}^T \; = \; 0,       
\nonumber \\      
& & \\      
{\rm Im} B_{+-}^T & = & \frac{- z k_T I_T}{\sqrt{z (1 - z)}}, \quad\quad {\rm Im} B_{-       
+}^T \; = \; \frac{(1 - z) k_T I_T}{\sqrt{z (1 - z)}}, \nonumber      
\eea      
for a photon of helicity +1, whereas for a longitudinal photon we have      
\bea      
\label{eq:a4}      
B_{++}^L & = & B_{--}^L \; = \; 0, \nonumber \\      
& & \\      
{\rm Im} B_{+-}^L & = & - {\rm Im} B_{-+}^L \; = \; \sqrt{
  \frac{z (1 - z) Q^2}{2}} \: I_L. \nonumber      
\eea      
The variable $z$ is the fraction of the photon's momentum carried by the quark, $k_T$ is the       
transverse momentum of the quark relative to the photon, $e_q$ is the charge (in units of $e$)       
and $m$ the mass of the quark; $\alpha = 1/137$.  The mass $M$ of the $q\bar{q}$ system       
satisfies      
\be      
\label{eq:a5}      
M^2 \; = \; \frac{m^2 + k_T^2}{z (1 - z)}.      
\ee      
The integrals over the transverse momentum $\pm \mbox{\boldmath
  $\ell$}_T$ of the exchanged gluons are~\cite{LMRT,MRT1}      
\bea      
\label{eq:a6}      
I_L (K^2) & = & \int \: \frac{d \ell_T^2}{\ell_T^4} \: \alpha_S (\ell_T^2) \: f (x, x^\prime,       
\ell_T^2)  \left (1 \: - \: \frac{K^2}{K_\ell^2} \right ), \\      
& & \nonumber \\      
\label{eq:a7}      
I_T (K^2) & = & \frac{K^2}{2} \int \frac{d \ell_T^2}{\ell_T^4} \: \alpha_S (\ell_T^2) \: f (x,       
x^\prime, \ell_T^2) \: \left [ \frac{1}{K^2} \: - \: \frac{1}{2k_T^2} \: + \: \frac{K^2 - 2k_T^2       
+ \ell_T^2}{2k_T^2 K_\ell^2} \right ],       
\eea      
where      
\bea      
\label{eq:a8}      
K^2 & = & z (1 - z) Q^2 \: + \: k_T^2 \: + \: m^2, \\      
& & \nonumber \\      
\label{eq:a9}      
K_\ell^2 & = & \sqrt{(K^2 + \ell_T^2)^2 \: - \: 4 k_T^2 \ell_T^2}.      
\eea      
The function $f (x, x^\prime, \ell_T^2)$ is the skewed unintegrated gluon distribution       
describing the lower part of Fig.~1.  The momentum fractions carried by the exchanged       
gluons satisfy \cite{MR}      
\be      
\label{eq:b9}      
\left ( x \: \simeq \: \frac{Q^2 + M^2}{W^2 + Q^2} \right ) \; \gg \; \left ( x^\prime \: \simeq \:       
\frac{\ell_T^2}{W^2 + Q^2} \right )      
\ee      
where $W$ is the $\gamma^* p$ centre-of-mass energy.      
      
In the strict leading $\log (1/x)$ approximation, it is enough to use the diagonal unintegrated       
distribution, as at each splitting we keep just the leading $\log (1/x)$ terms and neglect the       
corrections due to $x^\prime \ll x$.  In this limit      
\be      
\label{eq:c9}      
f (x, x^\prime, \ell_T^2) \; = \; f (x, \ell_T^2) \; = \; \left . \frac{\partial (xg (x,  
\mu^2))}{\partial \ln       
\mu^2} \right |_{\mu^2 = \ell_T^2}      
\ee      
and there is no difference between the diagonal and skewed distributions \cite{BL}.  This is       
the approximation which is conventionally used.      
      
Here we wish to take into account the skewed effect, but first we must extend the definition       
of the unintegrated gluon, (\ref{eq:c9}), beyond the leading $\log (1/x)$ approximation.        
Indeed it is easy to see that (\ref{eq:c9}) can only be true for sufficiently small $x$.  If $x$       
increases then $f$ calculated from (\ref{eq:c9}) would soon become negative due to the       
(negative) virtual contribution in the DGLAP evolution.  It was shown in Ref.~\cite{DDT}       
that the virtual corrections may be resummed via the Sudakov form factor and that the       
number of gluons with transverse momentum $\ell_T$ is      
\be      
\label{eq:d9}      
f (x, \ell_T^2) \; = \; \left . \frac{\partial [xg (x, q_0^2) \: T (q_0^2, \mu^2)]}{\partial \ln       
q_0^2} \right |_{q_0^2 = \ell_T^2},      
\ee      
where in the double $\log$ approximation (DLA)      
\be      
\label{eq:e9}      
T (q_0^2, \mu^2) \; = \; \exp \left [ \frac{- C_A \alpha_S (\mu^2)}{4 \pi} \: \ln^2 \:       
\frac{\mu^2}{q_0^2} \right ],      
\ee      
with scale $\mu^2 \sim (Q^2 + M^2)/4$.  $T$ is a survival probability.  It is the probability       
that the parent gluon does not emit gluons in the interval $q_0^2 < q_T^2 < \mu^2$.  From       
the formal point of view, the $T$ factor may be regarded as a next-to-leading order       
correction since the main contributions to the integrals (\ref{eq:a6}) and (\ref{eq:a7}) come       
from the region\footnote{If $\ell_T > \mu$ then we set $T = 1$ in (\ref{eq:d9}), consistent      
with the DLA.  It may occasionally happen (at the edge of phase space) that the inclusion of  
the $T$ factor in the DLA is not enough to ensure the positivity of $f (x, \ell_T^2)$, whereas  
the exact form of the $T$ factor would guarantee that $f > 0$.  Therefore we set $f = 0$ if it  
should happen that (\ref{eq:d9}) is negative.} $\ell_T^2 \lapproxeq \mu^2$.        
In general we find that the inclusion of $T$ has a small effect, essentially only ensuring the      
positivity of $f$ for $\Upsilon$ production which samples values of $x$ as large as $x \sim      
0.05$.      
      
The main effect of using the skewed (or off-diagonal) gluon distribution comes, within      
leading $\ln Q^2$       
kinematics, from the region where $x^\prime \ll x$, see (\ref{eq:b9}).  In this region the       
skewed gluon distribution $H_g (x, x^\prime)$ (integrated over $\ell_T$) is      
larger than the       
conventional diagonal distribution $H_g (x, x) = xg (x)$.  For small $x$, which is appropriate       
for vector meson production at HERA, the enhancement is generated entirely by off-diagonal       
evolution.  Moreover the ratio      
\be      
\label{eq:f9}      
R_g \; = \; \frac{H_g (x, x^\prime \ll x)}{H_g (x, x)}      
\ee      
can be determined unambiguously in terms of the known diagonal distribution \cite{SGMR}.        
It was shown that the enhancement $R_g$ depends on the effective power $\lambda (Q^2)$       
of the small $x$ behaviour of the gluon $xg \sim x^{-\lambda}$.  The result is \cite{SGMR}      
\be      
\label{eq:g9}      
R_g \; = \; \frac{2^{2 \lambda + 3}}{\sqrt{\pi}} \: \frac{\Gamma \left ( \lambda + \frac{5}{2}       
\right )}{\Gamma \left ( \lambda + 4 \right )}.      
\ee      
Note that the off-diagonal enhancement enters at leading order (in $\ln Q^2$) and increases       
with $Q^2$ (since $\lambda$ increases with $Q^2$).  Here we allow for the off-diagonal       
effect by multiplying the amplitudes (\ref{eq:a3}) and (\ref{eq:a4}), calculated\footnote{In       
the infrared region $\ell_T^2 < \ell_0^2$ we use the \lq\lq linear\rq\rq~approximation       
$\alpha_S (\ell_T^2) g (x, \ell_T^2) = (\ell_T^2/\ell_0^2) \alpha_S (\ell_0^2) g (x, \ell_0^2)$       
as described in \cite{LMRT}.  This linear approximation is reasonable since (i) it corresponds   
to a constant gluon-proton cross section at small scales $\ell_T < \ell_0$ and (ii) it matches   
well to the scale dependence of the phenomenological gluon distribution at low $\ell_T$.  We   
have checked that our results are stable to reasonable       
variations of $\ell_0^2$ about our default value of 1.5~GeV$^2$.} with diagonal gluons in       
(\ref{eq:a6}) and (\ref{eq:a7}), by the factor $R_g$.  We determine the effective power for       
each component amplitude separately, that is      
\be      
\label{eq:h9}      
\lambda \; = \; \frac{\partial \log B_{ii^\prime}}{\partial \log (1/x)}.      
\ee      
      
The full NLO corrections for the diffractive process are not known yet, and so we       
approximate them by a ${\cal K}$ factor \cite{LMRT,MRT1}.  Following \cite{LMRT}, the       
main ($\pi^2$ enhanced) part of the ${\cal K}$ factor is of the order of $(\pi^2 C_F       
\alpha_S/\pi)$, where $C_F = 4/3$.  It comes from the $i\pi$ terms in the double logarithmic       
Sudakov form factor $\exp [-C_F (\alpha_S/4 \pi) \ln^2 (-M^2)]$ where $\ln (-M^2) = \ln       
M^2 + i\pi$.  Thus we multiply the amplitudes by the factor \cite{LMRT}     
\be      
\label{eq:K}      
{\cal K} \; = \; \exp (\pi C_F \alpha_S/2).      
\ee      
But we still have the ambiguity of the choice of the scale of $\alpha_S$.  We show results for       
two choices of scale:  $\mu^2 = K^2$ and our default value $\mu^2 = 2K^2$, where $K^2$ is   
given by       
(\ref{eq:a8}).      
      
To include the contribution from the real part of the amplitude we use the signature factor      
\be      
\label{eq:j9}      
{\cal S}^{(+)} \; = \; i \: + \: \tan (\pi \lambda/2)      
\ee      
for positive signature exchange.  This is a simple way of implementing the dispersion relation       
result.  It gives      
\be      
\label{eq:k9}      
{\rm Re} B_{ii^\prime} \; = \; \tan (\pi \lambda/2) \: {\rm Im} B_{ii^\prime},      
\ee      
where $\lambda$ is given by (\ref{eq:h9}).  The inclusion of the real
part enhances the cross section of $\rho$ production by 14 to 19\% in
the range where we compare to data, $J/\psi$ production by 18 to
25\%, and $\Upsilon$ by about 30\%, where the bigger effect always
occurs at higher $Q^2$.
      
So far we have calculated the cross section $d^2 \sigma/dM^2 dt$ for diffractive $q\bar{q}$       
production at $t = 0$.  To determine $d\sigma/dM^2$ we integrate the form $\exp (- bt)$ over       
$t$, with \cite{RSS}      
\be      
\label{eq:l9}      
b (Q^2) \; = \; \frac{4}{(\langle t \rangle + 0.71~{\rm GeV}^2)} \: + \: \frac{2}{Q^2 + M^2 +      
\langle t \rangle} \: + \: 2       
\alpha_{\funp}^\prime \: \ln \left ( \frac{W^2 M^2}{(Q^2 + M^2)^2} \right ),     
\ee      
where $\langle t \rangle$ is the average value of $t$.  (Here we set
$\langle t \rangle = 0$.)  This form, with $\alpha_\funp^\prime =      
0.15~{\rm GeV}^{-2}$, successfully reproduces the       
$t$ behaviour of diffractive $\rho$ meson leptoproduction data as a function of $Q^2$ and       
$M^2 \simeq M_V^2$.  It is motivated by the additive quark model, together with a form       
factor given by $F_V (t) \; = \; M^2/(M^2 - t)$, see also \cite{HKK}.  We also use the      
phenomenological expression (\ref{eq:l9}) for diffractive $J/\psi$ production even though the      
measured slopes appear, at present, to be about 2~GeV$^{-2}$ or 30\% less.  Using the      
observed values of $b$ would lead to an overall increase in the $J/\psi$ cross section of about   
30\%, well within the present uncertainties in the theoretical normalization.     
      
To determine the cross section for $\gamma^* p \rightarrow Vp$ from that for diffractive       
$q\bar{q}$ production, we project out the $J^P = 1^-$ state in the $q\bar{q}$ rest frame.        
However the helicity amplitudes $B_{ii^\prime}$ are defined in the target proton rest frame,       
and helicity is not conserved by Lorentz transformations for the heavy quark states.  So to  
obtain the helicity amplitudes $A_{jj^\prime}$ in the $q\bar{q}$ rest frame for $V = J/\psi$  
and $\Upsilon$, we must perform a Lorentz boost and use      
\be      
\label{eq:i9}      
A_{jj^\prime} \; = \; \sum_{i, i^\prime} \: c_{ij} \: c_{j^\prime i^\prime} \: B_{ii^\prime},      
\ee      
where the known coefficients $c_{ij}$ are given in Ref.~\cite{MRT3}. Finally we integrate      
the cross section $d \sigma/dM^2$ for $J = 1^- \: q\bar{q}$ production over an appropriate      
interval $\Delta M^2$ covering the vector meson resonance.  Clearly this, together with the      
${\cal K}$ factor of (\ref{eq:K}), introduces an overall normalization uncertainty.  However      
here we are interested in the $Q^2$ dependence of $\sigma (\gamma^* p \rightarrow Vp)$      
and the properties of the ratio $\sigma_L/\sigma_T$, rather than the normalization.      
      
\begin{figure}[htb]
\vspace{-0.5cm}
\begin{center}
\leavevmode
\epsfxsize=12.0cm
\epsffile[70 260 480 560]{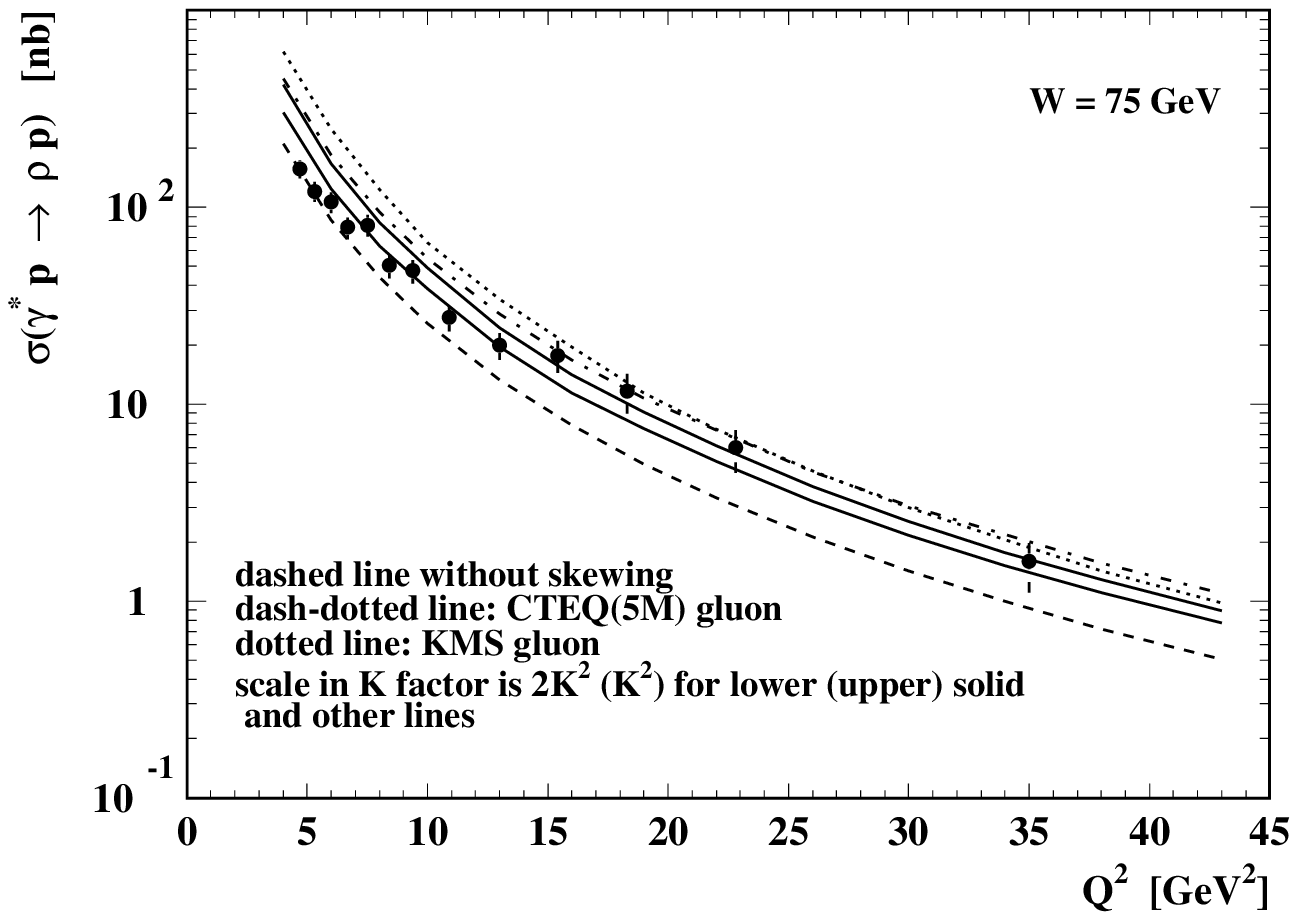}\\
\vspace{-0.6cm}
\leavevmode
\epsfxsize=12.0cm
\epsffile[70 260 480 560]{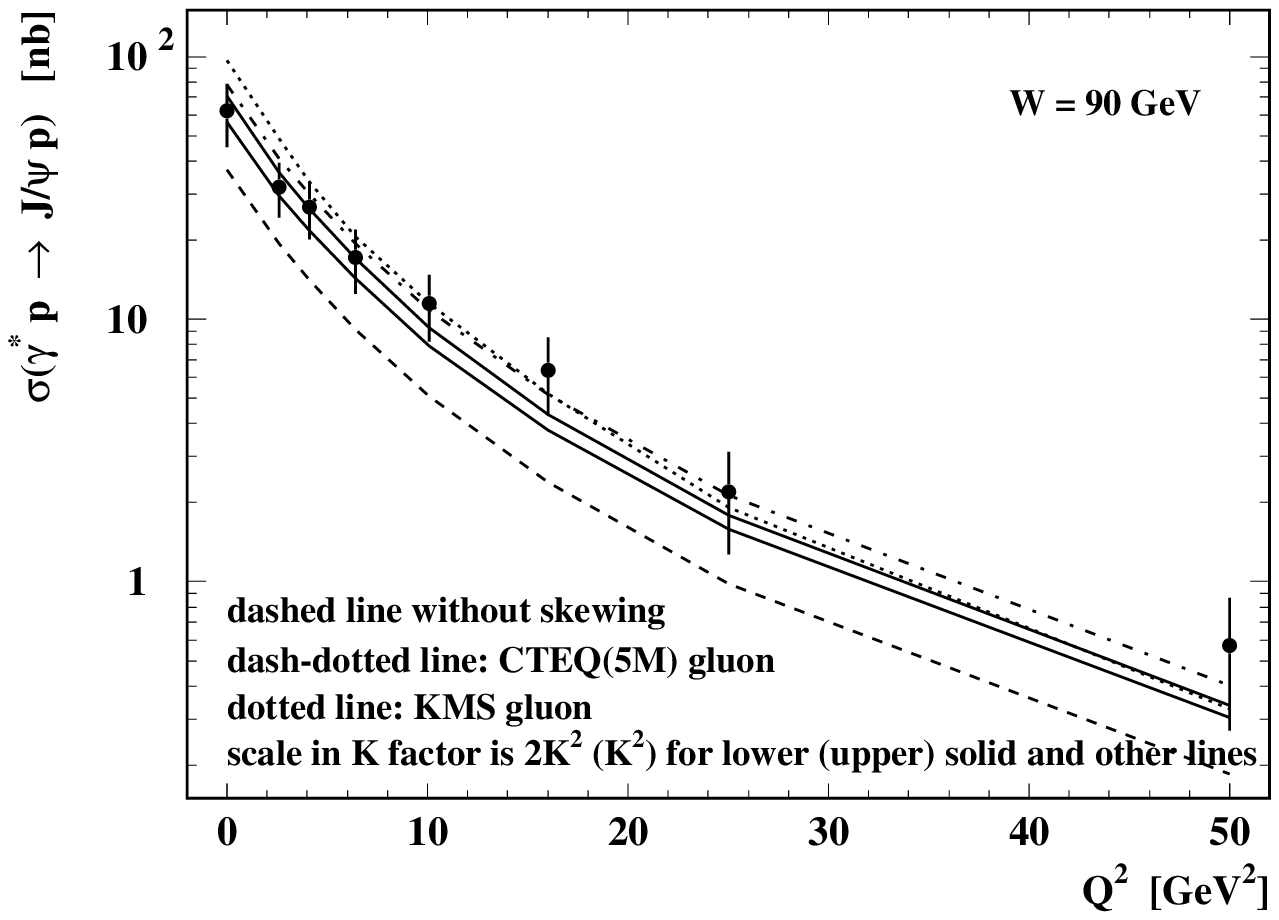}
\end{center}
\vspace{-0.6cm}
\caption[]{\label{fig2} The QCD predictions for the $Q^2$ dependence of
  the cross sections for $\gamma^* p \rightarrow \rho p$ (upper plot, $W =
  75$ GeV) and $\gamma^* p \rightarrow J/\psi \, p$ (lower plot, $W =
  90$ GeV) compared with the HERA data \cite{H2,H3}.  The continuous
  curves are obtained using the MRST(99) gluon \cite{MRST}.  For the
  lower curve the default value $2K^2$ is chosen for the scale of
  $\alpha_S$ in the ${\cal K}$ factor, whereas for the upper curve the
  scale $K^2$ was used.  The dash-dotted (dotted) curves show the
  results if the CTEQ(5M) \cite{CTEQ5} (KMS \cite{KMS}) gluon are
  used.  The dashed curves show our results using the MRST(99) gluon
  and default parameters but without the effect of skewing.  All
  predictions contain contributions from the real part of the
  amplitude as discussed in the text.  The data point for $J/\psi$
  photoproduction in the lower plot is interpolated between H1 data
  for different values of $W$ \cite{H4} and agrees well with the ZEUS
  result~\cite{Z3}.}
\end{figure}
The QCD predictions for $\rho$, $J/\psi$ and $\Upsilon$
production\footnote{The $J/\psi$ production amplitudes were calculated
  for a charm quark mass $m_c = 1.4$~GeV, whereas for $\Upsilon$
  production we take the $b$ quark mass $m_b = 4.6$ GeV.} are compared with   
HERA data in Figs.~2--4.  
Fig.~2 shows the predictions obtained from
using three different   
recent gluon distributions:  MRST \cite{MRST}, CTEQ \cite{CTEQ5} and KMS   
\cite{KMS}.           
All fit the $F_2$ structure function data well.  The first two are obtained from          
conventional NLO DGLAP analyses, while the KMS analysis is in terms of an          
unintegrated gluon distribution which satisfies a unified BFKL/DGLAP equation with          
subleading $\ln (1/x)$ contributions.  In each case the scale $\mu^2$ in $\alpha_S$ of the      
${\cal K}$ factor is taken to be $2K^2$.  The lower continuous curves
in Fig.~2 correspond to the \lq      
default\rq~prediction obtained using MRST99 partons.  In both plots
the upper continuous      
curve corresponds to taking $\mu^2 = K^2$.  The dashed curve is the default prediction using      
the diagonal gluon, and so comparison with the lower continuous curve shows the      
enhancement due to off-diagonal effects.  At the larger values of $Q^2$ the enhancement is      
about 55\% for $\rho$ production and 70\% for $J/\psi$ production.  


Our approach is infrared finite.  However there are non-negligible
contributions from the region of low gluon transverse momenta $\ell_T
< \ell_0$.  Fortunately the predictions based on conventional DGLAP
partons (MRST, CTEQ) are rather insensitive to the choice of the value
of $\ell_0$.  Nevertheless to check the infrared sensitivity of the
predictions we also use a gluon obtained from a unified BFKL/DGLAP
analysis of the deep inelastic data \cite{KMS}.\footnote{In \cite{KMS}
  the value of the unintegrated gluon is determined down to $\ell_T =
  k_0 =1$ GeV. Below $k_0$ we use the linear approximation as was described 
     in footnote 4 above, with $xg(x,k_0) = 1.57\,(1-x)^{2.5}$.}  We
   would expect some difference since the
latter (Reggeised) gluon embodies a higher twist component originating
from the BFKL evolution, which may be important at low scales.  Indeed
we see from the dotted curves in Fig.~2 that the cross sections are
considerably larger than the DGLAP-based predictions particularly at
low values of $Q^2$.  This demonstrates the need to better understand
the role of higher twist (and power) corrections in parton analyses.
Diffractive vector meson production is clearly a good process in which
to further investigate these effects.

In the parton-hadron duality approach we have a common mechanism for
the description of all vector meson production processes, $\gamma^* p
\rightarrow Vp$, governed by the average hard scale $\langle K^2
\rangle$, where    
\begin{equation}
\label{eq:a22}
K^2 \; = \; z (1 - z) (Q^2 + M_V^2).
\end{equation}
Therefore it is informative to plot all the observed cross sections,
in a given $W$ domain, as a function of $Q^2 + M_V^2$ on the same
plot, after allowing for the different photon-quark couplings $e_q$
(that is $\rho : J/\psi : \Upsilon = 9 : 8 : 2$) and the different
energies $W$ of the data.  The result is compared with the $\rho$
prediction in Fig.~3.  
\begin{figure}[htb]
\vspace{-0.5cm}
\begin{center}
\leavevmode
\epsfxsize=14.0cm
\epsffile[0. 10. 440. 530.]{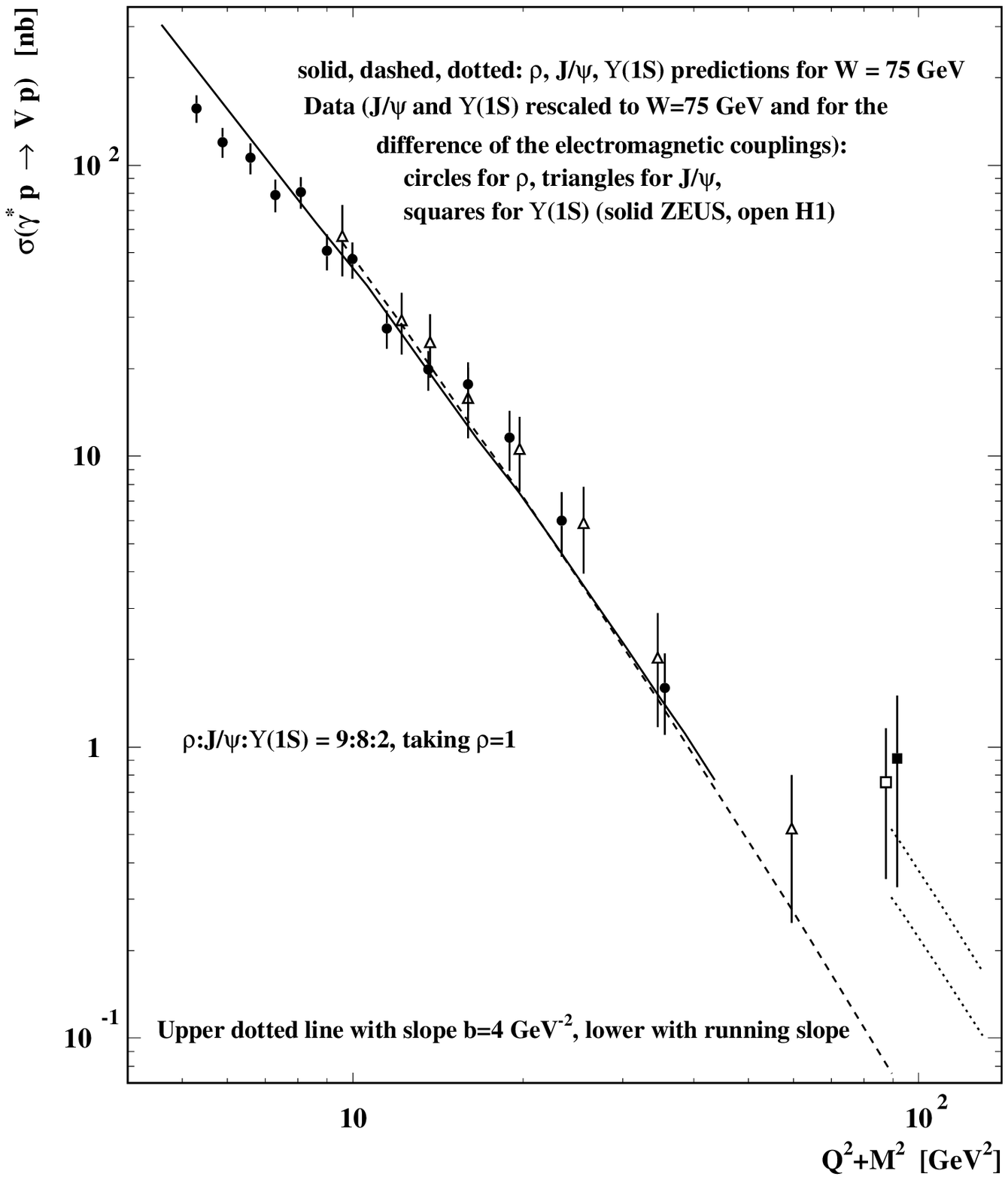}
\end{center}
\vspace{-0.5cm}
\caption[]{\label{fig3} The data \cite{H2,H3,H4,UH1,UZ} for the $\gamma^* p
  \rightarrow Vp$ cross sections with $V = \rho$ (circles), $J/\psi$
  (triangles) and $\Upsilon(1S)$ (squares: solid ZEUS, open H1, both
  slightly displaced from $Q^2 = 0$ for readability) plotted versus
  $Q^2 + M_V^2$.  The QCD predictions (with standard parameters as
  described in the text) are shown for comparison as continuous, dashed
  and dotted lines, respectively.  The $J/\psi$ and $\Upsilon$ data
  (and errors) are corrected for (i) the different photon-quark
  couplings by multiplying the $J/\psi$ and $\Upsilon$ measurements by
  9/8 and 9/2 respectively, (ii) the different $W$ values according to
  the QCD predicted energy behaviour $\sigma (J/\psi) \sim W^{1.1}$
  (in agreement with the experimental measurements from \cite{H3}) and
  $\sigma (\Upsilon) \sim W^{1.3}$.  The upper dotted curve is
  obtained using a fixed slope parameter $b = 4$ GeV$^{-2}$, whereas
  the lower curve contains the slope as given in Eq.~(\ref{eq:l9}).}
\end{figure}
The fact that the measured cross sections approximately lie on a common curve,  
demonstrates the universality inherent in the perturbative QCD description.  Some departure 
from universality may arise from the different measured $t$ slopes, from the flavour 
symmetry breaking of the $q\bar{q} \rightarrow V$ transition\footnote{In the non-relativistic  
approximation the $q\bar{q} \rightarrow V$ vertices are proportional to the meson wave  
functions evaluated at the origin, which may differ according to the mass of the meson.  In  
our approach this is replaced by the different $z$ intervals sampled by the relativistic $\rho$  
system as compared to the more non-relativistic $J/\psi$ and
$\Upsilon$ systems and by possible different choices of the mass
interval $\Delta M$ covering the resonance peaks.  Here the  
same value $\Delta M = 200$~MeV was used for $\rho$ and $J/\psi$
production. For the $\Upsilon$ the interval $M = 8.9 \ldots
10.9$ GeV was chosen to predict $\Upsilon(1S,\,2S,\,3S)$ production in
accordance with the experimental analysis \cite{UZ}, and the resulting
cross sections where divided by 1.7 to get the predictions for
$\Upsilon(1S)$, see \cite{MRT3}.}, and from  
comparing data with different average $W$ values.     
     
\begin{figure}[htb]
\vspace{-0.5cm}
\begin{center}
\leavevmode
\epsfxsize=12.cm
\epsffile[80. 270. 470. 560.]{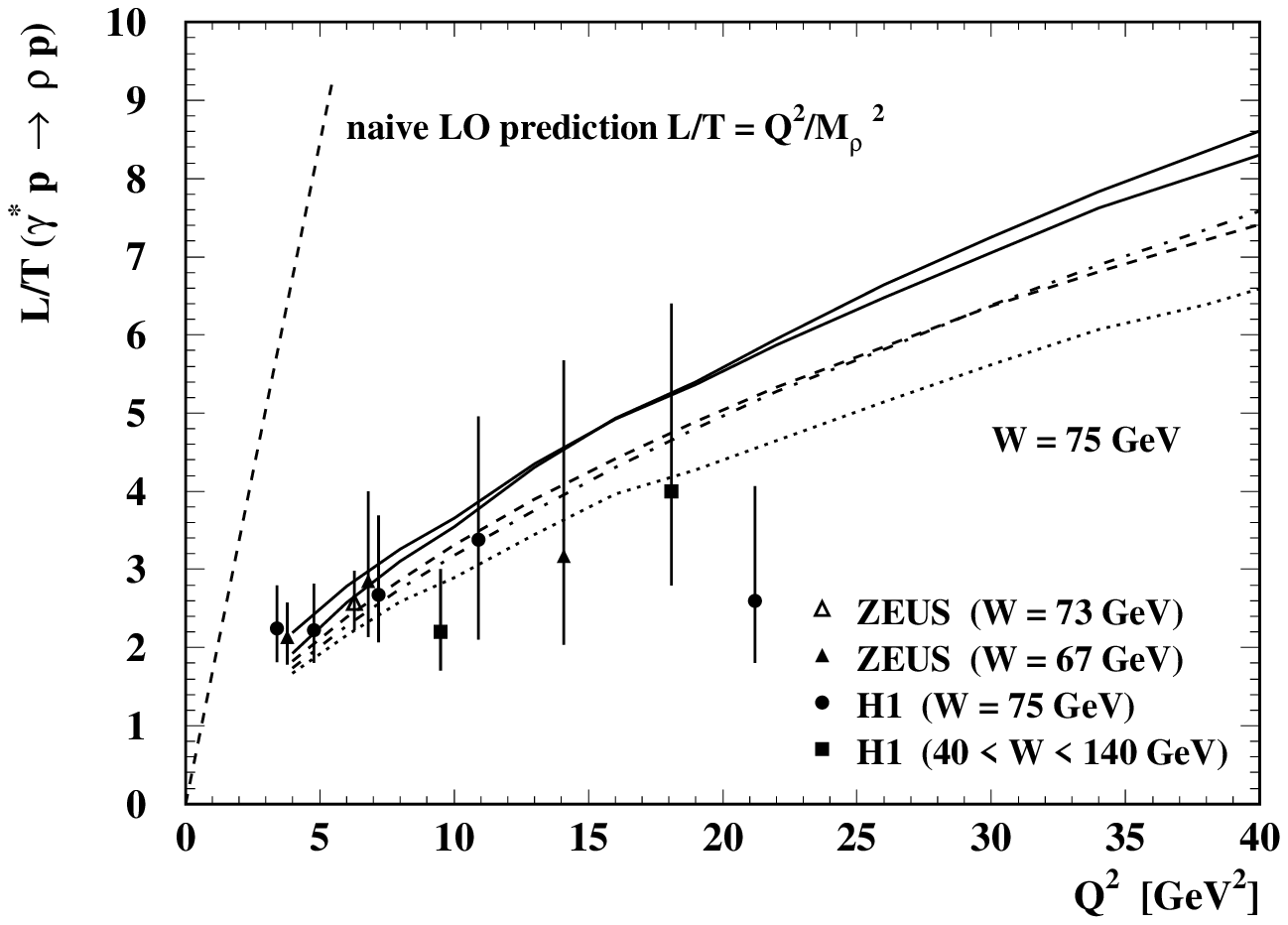}\\
\vspace{-0.5cm}
\leavevmode
\epsfxsize=12.cm
\epsffile[80. 270. 470. 560.]{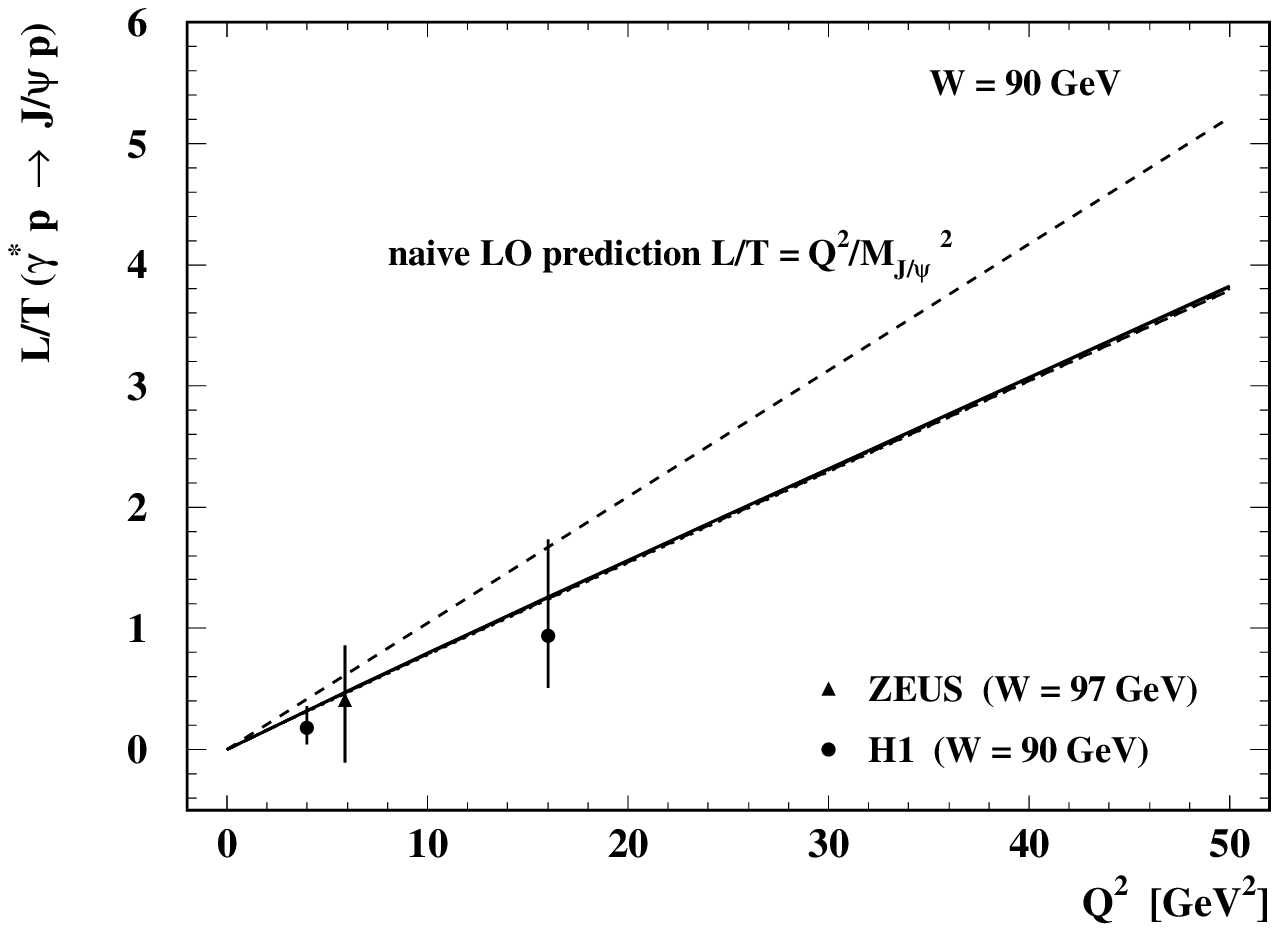}
\end{center}
\vspace{-0.5cm}
\caption[]{\label{fig4} The upper plot shows the QCD predictions for the $Q^2$
  dependence of $\sigma_L/\sigma_T$ for $\rho$ electroproduction (at
  $W = 75$ GeV) compared with HERA data \cite{H2,H1,Z1,Z2}, partially
  at slightly different (average) values of $W$ as indicated on the
  plot. The ZEUS measurement displayed by the open triangle is the one
  obtained by relaxing the $s$-channel helicity conservation
  condition, see~\cite{Z2}.  The
  different linestyles for the different gluons are chosen as in
  Fig.~2.  Here the steeper continuous curve corresponds to the
  standard choice of $2K^2$ as scale of $\alpha_S$ in the ${\cal K}$
  factor, the less steeper one to $K^2$.  Also displayed is the naive
  expectation $\sigma_L/\sigma_T = Q^2/M_{\rho}^2$ (steep dashed
  line).  The lower plot shows $\sigma_L/\sigma_T$ for
  $J/\psi$ production (at $W = 90$ GeV) compared to data from \cite{H3,Z1}.} 
\end{figure}
\begin{figure}[htb]
\vspace{-0.5cm}
\begin{center}
\leavevmode
\epsfxsize=12.cm
\epsffile[80. 270. 470. 560.]{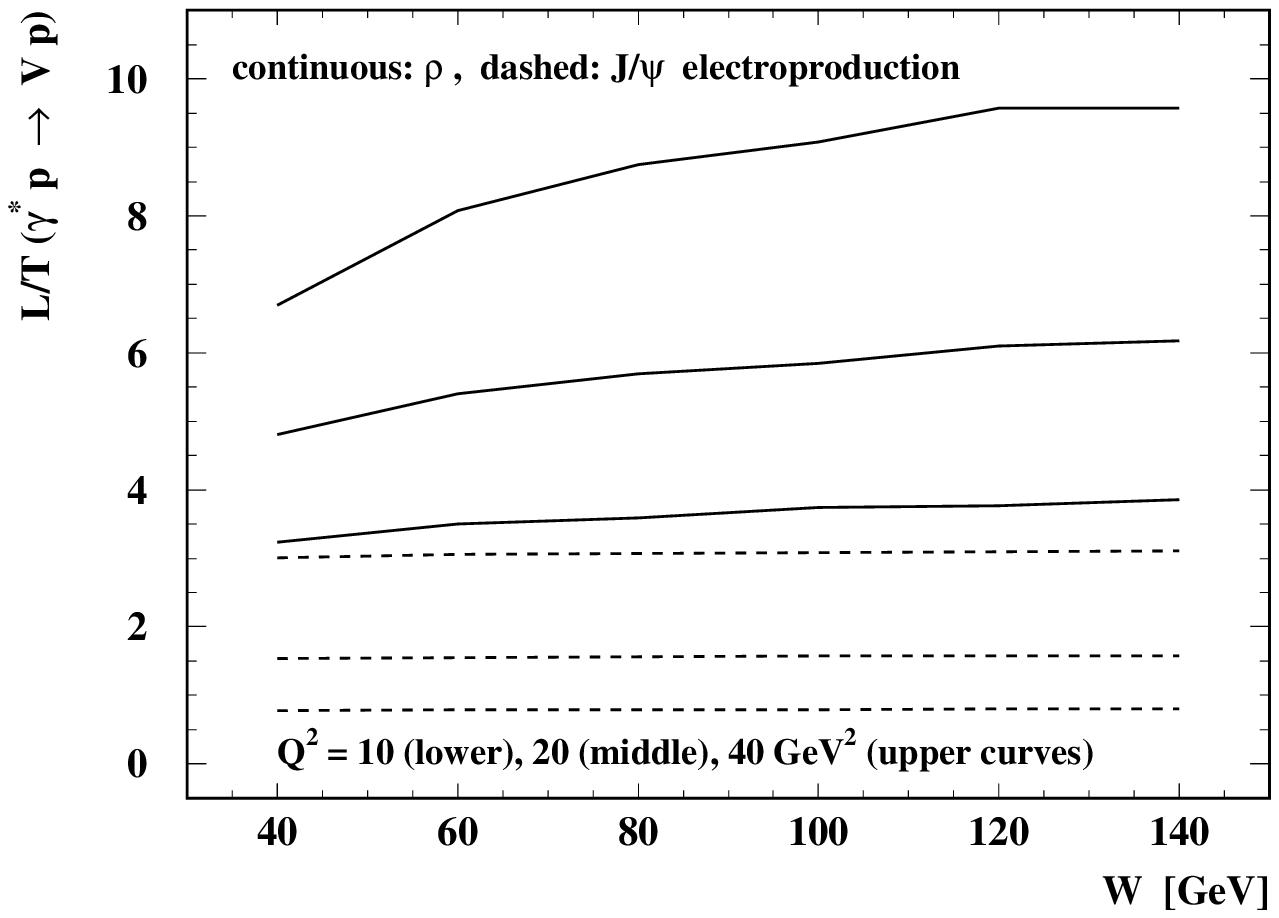}
\end{center}
\vspace{-0.5cm}
\caption[]{\label{fig5} The $W$ behaviour of $\sigma_L/\sigma_T$ for
  fixed values of $Q^2$ for both $\rho$ electroproduction (continuous
  curves) and $J/\psi$ electroproduction (dashed curves), obtained
  with our default parameters and the MRST(99) gluon.}
\end{figure}

Figs.~4, 5 show the QCD predictions for $\sigma_L/\sigma_T$.  The upper
(lower) plot of Fig.~4 compare the ratio for $\rho$ ($J/\psi$)
electroproduction with the recent HERA data as a function of $Q^2$ at
fixed energy $W = 75$ GeV ($W = 90$ GeV), whereas Fig.~5 shows
the $W$ dependence for fixed values of $Q^2$ for both $\rho$ and
$J/\psi$ production.  Recall that in Ref.~\cite{BROD} it was pointed out      
that only $\sigma_L$ is calculable in perturbative QCD; the calculation of $\sigma_T (\rho)$      
using the leading twist $\rho$ meson wave function is infrared divergent.  We must therefore      
explain how the $\sigma_L/\sigma_T$ curves can be obtained?  In the   
parton-hadron duality approach, with the $J^P = 1^-$ projection of the $q\bar{q}$ system,   
the integral over the quark $k_T$ is of logarithmic form \cite{MRT1}
(like in a usual DIS amplitude).  So the corresponding Feynman graphs
have (at leading order) a pure ladder structure with strong $k_T$
ordering along the ladder.  The factorization   
theorem is therefore valid for $\sigma_T$, as well as $\sigma_L$.  After convolution with the   
gluon distribution, the logarithmic behaviour effectively enhances the transverse amplitude by   
a factor $1/\gamma$, so $\sigma_T \sim 1/\gamma^2$ as in (\ref{eq:aa}).  The decrease of   
$\gamma$ with increasing $Q^2$ masks the naive $Q^2/M^2$ expectation for the $Q^2$   
behaviour of $\sigma_L/\sigma_T$.    
     
It is interesting to compare the predictions for the $W$ behaviour of $\sigma_L/\sigma_T$      
for $J/\psi$ production with those for $\rho$ production, shown
respectively by the dashed and continuous curves in Fig.~5.  For
$\rho$ production we are in a relativistic $q\bar{q}$      
situation where $z$ covers an extensive part of the (0,1) interval allowing the $1/\gamma$      
 behaviour to develop.  The growth of $\sigma_L/\sigma_T$ with $W$ reflects the rise of   
$\gamma$ with $1/x$.  On the other hand $J/\psi$ production is nearer the non-relativistic      
limit where $z = 1/2$ and $\sigma_L/\sigma_T = Q^2/M^2$ apply, and hence the ratio   
$\sigma_L/\sigma_T$ is almost independent of $W$.     
     
To gain physical insight we have discussed the results in terms of an effective gluon   
anomalous dimension $\gamma$ and      
the simplified formula (\ref{eq:aa}).  In the actual computations we use, of course, the      
explicit unintegrated (skewed) gluon distributions and perform the full
integrations over $k_T$ and $\ell_T$, or related variables.     
     
In summary we have shown that a perturbative QCD parton-hadron duality approach is able      
to describe all the main features of the $\sigma_L$, and even the $\sigma_T$, data for      
diffractive vector meson production $\gamma^* p \rightarrow Vp$, {\it provided} that there      
is a sufficiently hard scale (that is provided $Q^2 + M^2$ is greater than about 5~GeV$^2$).   
We emphasize that the approach is infrared convergent.  There are
non-negligible contributions at low gluon transverse momentum
$\ell_T$, but the perturbative gluon form      
matches well on to the linear $\ell_T^2$ form for $\ell_T < \ell_0$ making the predictions      
rather insensitive to the choice of $\ell_0$.  The effects of using skewed gluons are fully  
included in the QCD calculations.  The skewed distribution is completely determined by the  
conventional (diagonal) gluon distribution, and is found to enhance the $\rho$ and $J/\psi$  
cross sections by about 55\% and 70\% respectively at the largest observed values of $Q^2$.   
For $\rho$ production the use of the skewed gluon distribution predicts a flatter $Q^2$  
dependence, compatible with the recent data, see Fig.~2.  We conclude
that the data for diffractive vector meson production processes at
HERA offer a particularly sensitive probe of the properties of the
gluon distribution of the proton.  The $M$, $Q^2$, $t$ dependences and
the spin properties can be measured with increased precision and all 
constrain the behaviour of the gluon.\\ 
     
\noindent {\large \bf Acknowledgements}     
     
One of us (MGR) thanks the Royal Society and the Russian Fund for Fundamental Research   
(98-02-17629) for support.  Part of this work was carried out while TT was at DESY.  The  
work was also supported by the EU Framework TMR programme, contract  
FMRX-CT98-0194.     
We thank Anna Stasto for valuable information.

\end{document}